\documentstyle[twoside]{article}
\def\SU{{\rm SU}}
\def\SO{{\rm SO}}
\def\Tr{\mathop{\rm Tr}}
\def\diag{\mathop{\rm diag}}

\makeatletter
\newcounter{alphaequation}[equation]
\def\thealphaequation{\theequation\alph{alphaequation}}
\def\eqnsystem#1{
\def\@eqnnum{{\rm (\thealphaequation)}}
\def\@@eqncr{\let\@tempa\relax
\ifcase\@eqcnt \def\@tempa{& & &}
\or \def\@tempa{& &}\or \def\@tempa{&}\fi\@tempa
\if@eqnsw\@eqnnum\refstepcounter{alphaequation}\fi
\global\@eqnswtrue\global\@eqcnt=0\cr}
\refstepcounter{equation}
\let\@currentlabel\theequation
\def\@tempb{#1}
\ifx\@tempb\empty\else\label{#1}\fi
\refstepcounter{alphaequation}
\let\@currentlabel\thealphaequation
\global\@eqnswtrue\global\@eqcnt=0
\tabskip\@centering\let\\=\@eqncr
$$\halign to \displaywidth\bgroup
  \@eqnsel\hskip\@centering
  $\displaystyle\tabskip\z@{##}$&\global\@eqcnt\@ne
  \hskip2\arraycolsep\hfil${##}$\hfil&
  \global\@eqcnt\tw@\hskip2\arraycolsep
  $\displaystyle\tabskip\z@{##}$\hfil
  \tabskip\@centering&\llap{##}\tabskip\z@\cr}

\def\endeqnsystem{\@@eqncr\egroup$$\global\@ignoretrue}
\makeatother

\begin{document}\large
\hfill\vbox{\baselineskip12pt
            \hbox{\bf IFUP -- TH 20/94}
            \hbox{\bf hep-ph/9404278}
            \hbox{April 1994}}
\vspace{7mm}
\begin{center}\vglue 0.6cm{\Large\bf\vglue 10pt
   Strings versus supersymmetric GUTs:\\ \vglue 3pt
   can they be reconciled?    \\}
\vglue 1.0cm
{\large\bf  Riccardo Barbieri, Gia Dvali$^*$ and Alessandro Strumia\\[4mm] }
\baselineskip=13pt {\em Dipartimento di Fisica, Universit\`a di Pisa and
\\[2mm] }
\baselineskip=12pt {\em INFN, Sezione di Pisa, I-56126 Pisa, Italy\\[6mm]}
\vfill

{\large\bf Abstract}
\end{center}

\vglue 0.3cm{\rightskip=3pc \leftskip=3pc \tenrm\baselineskip=12pt
\noindent\large
We describe a class of supersymmetric unified models
with the following properties:
i) the full breaking of the gauge group is achieved by Higgs fields in the
fundamental representation;
ii) the correct unification of the strong and electroweak coupling constants
is obtained without the need of any intermediate scale;
iii) the problems of the doublet-triplet splitting and of the
proton decay at dimension-5 level may receive a natural solution.
The models, other than being interesting unified field theories {\em per se\/},
may constitute examples of string-derivable GUTs.}

\vfill\vfill\normalsize\footnoterule~\\[0.5mm]
\noindent
{$\ast$~~Permanent address:\em{}
Institute of Physics, Georgian Academy of Sciences, 380077 Tbilisi, Georgia.}

\thispagestyle{empty}
\newpage~
\setcounter{page}{1}

\normalsize
\section{}
In this note we describe a class of supersymmetric
models based on the gauge group $G\otimes G$, with $G\supseteq\SU(5)$,
with the following properties:
\begin{itemize}
\item[i)] the full breaking of the gauge group is achieved by Higgs fields
in the fundamental representation;
\item[ii)] the correct unification of the strong and electroweak gauge
couplings is obtained without the need of any intermediate scale;
\item[iv)] the problems of the doublet-triplet splitting and of the
proton decay at dimension-5 level may receive a natural solution.
\end{itemize}
The models, other than being interesting unified field theories {\em per se\/},
may constitute examples of string-derivable GUTs.
We have in mind the difficulty to obtain, in a string theory context, gauge
models with adjoint or higher representations in their spectra~\cite{spectra}.
Our focus, on the other hand, is on the comparison of standard GUTs~\cite{GUTs}
versus ununified string models~\cite{noGUTstring} in their prediction
of the gauge couplings at low energy. In this respect,
ununified string models with conventional $k_i$-factors, although in
principle more predictive, are not as successful as standard GUTs,
unless large string threshold corrections are invoked~\cite{strThreshold}.
Such corrections seem indeed to be there.
The question is however: why should these corrections maintain the relation
between the couplings characteristic of the Grand Unified symmetry,
if such a symmetry is not actually realized?~\cite{strUnif}
Needless to say, flipped $\SU(5)\otimes{\rm U}(1)$~\cite{SU5U1} does not
differ in this respect from $\SU(3)\otimes\SU(2)\otimes{\rm U}(1)$ or any
ununified group.

\section{}
For concreteness we discuss an $\SU(5)'\otimes\SU(5)''$ model.
The trivial extension to larger groups, in particular to
$\SO(10)\otimes\SO(10)$ or $\SU(6)\otimes\SU(6)$,
will be briefly described later on.
The full breaking of the gauge group to the standard
$\SU(3)\otimes\SU(2)\otimes{\rm U}(1)$ is achieved by the vacuum
expectation values (vev) of one, or more, multiplets
\begin{equation}\label{eq:Z}
Z_{a'i}^{a''}=(5,\bar 5)_i,\qquad
\bar{Z}^{a'}_{a''i}=(\bar5,5)_i
\end{equation}
with $\SU(5)'$ and
$\SU(5)''$ indices, $a'$ and $a''$,
in the fundamental respective representations\footnote{From the point of view
of the present paper, $Z,\bar{Z}$ multiplets transforming as $(5,5)$ and
$(\bar5,\bar5)$, rather than $(5,\bar 5)$ and $(\bar5,5)$, with a consequent
change in the representation of the matter multiplets, are equally apt to
their purpose. The two cases, however, may not be equivalent from the point of
view of a more fundamental theory.}.
With a generic superpotential $W(Z_i,\bar{Z}_i)$,  it is easy to
show that the supersymmetric potential in the scalar components of the
superfields $Z_i,\bar{Z}_i$ has a supersymmetric minimum
for $Z=\bar{Z}$ of the following possible forms
\begin{eqnsystem}{Z1234}
Z_1 &=& V_1\cdot\diag(1,1,1,1,1),\label{eq:Z1}\\
Z_2 &=& V_2\cdot\diag(0,0,0,1,1),\label{eq:Z2}\\
Z_3 &=& V_3\cdot\diag(1,1,1,0,0),\label{eq:Z3}\\
Z_4 &=& V_4\cdot\diag(1,1,1,x,x),\label{eq:Z4}\qquad x\neq1.
\end{eqnsystem}
In general it is sufficient that the superpotential be dependent at least
upon two invariants, like $\Tr(Z\bar{Z})$ and $\Tr(Z\bar{Z}Z\bar{Z})$.
Various $Z_i,\bar{Z}_i$ fields may be coupled to each other in the
superpotential, as needed to avoid unwanted massless particles, and still
have different orientations of their vevs, as in eq.s~(\ref{Z1234}).

These vevs lead respectively to the breakings of the gauge group
$\SU(5)'\otimes\SU(5)''$ down to
\begin{eqnsystem}{eq:SU123}
&&\SU(5)\\
&&\SU(3)'\otimes\SU(3)''\otimes\SU(2)\otimes{\rm U}(1)\\
&&\SU(3)\otimes\SU(2)'\otimes\SU(2)''\otimes{\rm U}(1)\\
&&\SU(3)\otimes\SU(2)\otimes{\rm U}(1)
\end{eqnsystem}
where the unprimed factors correspond to obvious diagonal
subgroups\footnote{The possibility to break a group $G\otimes G$ in an
interesting way by means of Higgs multiplets in the fundamental
representation is pointed out in ref.~\cite{fundHiggs}.}.

Any pair of the first three vevs, as~(\ref{eq:Z4}) alone, give rise to the
usual low energy group.
Furthermore, irrespective of the values of the $\SU(5)'$ and $\SU(5)''$
couplings, $g'$ and $g''$, the standard gauge couplings $g_3,g_2,g_1$ are
unified at a common scale $M$, as in standard SU(5), in any of the following
cases (among others):
\begin{itemize}
\item[i)] $V_1\raisebox{-.4ex}{\rlap{$\sim$}} \raisebox{.4ex}{$>$}
V_2\approx V_3\approx M$ (no $V_4$);
\item[ii)] $V_2\approx V_3\approx M$ (no $V_1,V_4$);
\item[iii)] $V_4\approx M$ (no $V_1,V_2,V_3$).
\end{itemize}
In case i) the theory is actually indistinguishable from simple SU(5) up to the
scale $V_1$, which may be arbitrarily high.
Notice that in all cases the low energy group lives in the diagonal SU(5).
We find this essential to achieve the desired unification of couplings.
Had we considered either SU(3) or SU(2) embedded in $G'$ or $G''$, with the
symmetric group factor broken at high energy, we would have not obtained
the correct boundary condition on the gauge couplings even with $g'=g''$.
In turn, this is what prevents the consideration of models fully
symmetric under interchange of the simple group factors $G'$ and $G''$,
since one does not want a doubling of the light matter particles
with symmetric Yukawa couplings\footnote{An interesting possible way out is
offered by theories based on the gauge group $G\otimes G\otimes G$.}.

\section{}

The matter multiplets are taken to transform as $\bar{5}\oplus10$
representations of either SU(5) factors.
It is in fact possible that different families transform under different
SU(5)'s, leading to a possibly interesting asymmetry between them.
For simplicity we take here
\begin{equation}
f_i=(\bar{5}\oplus10,1)_i\equiv(\bar{5}'\oplus10')_i,\qquad i=1,2,3.
\end{equation}
{}From the point of view of the diagonal SU(5), the light Higgs doublets
must live in $5\oplus\bar{5}$ representations.
Also in this case, therefore, one has several possible choices for the
transformation properties of the corresponding multiplets under the
full $\SU(5)'\otimes\SU(5)''$ group.
An interesting possibility, also taking into account the anomaly cancellation
requirement, would be to take
$$ H'=(\bar{5}\oplus 10,1),\qquad
\bar{H}''=(1,5\oplus \overline{10}).$$
In the following, however, we shall stick to the choice
$$
H' =(5,1),\qquad \bar{H}' =(\bar5,1), \qquad
H''=(1,5),\qquad \bar{H}''=(1,\bar5).
$$
with a doubling $H',\bar{H}'\to H'',\bar{H}''$ that may or may not
be necessary.

The presence of $Z$-multiplets with the vevs~(\ref{eq:Z2}), (\ref{eq:Z3})
suggests simple ways to overcome the difficulties of usual GUTs associated
with the doublet-triplet splitting and with the possible dimension-5
operators mediating the proton decay~\cite{pDecCalc}.
On one side, the heavy mass for the triplet fields may arise from couplings
of the form
$$H' Z_3 \bar{H}'',\qquad \bar{H}'\bar{Z}_3 H''.$$
On the other side, the same triplets may even be decoupled from the light
generation if the required Higgs coupling is obtained through the
non-renormalizable operators\footnote{A mechanism of this type to suppress
the proton decay has been suggested
in an SO(10) context in ref.~\cite{NoPdecay}.}
$$\frac{1}{M}\bar{5}'10'\bar{Z}_2\bar{H}'',\qquad
\frac{1}{M}10'~10' Z_2 H''.$$
At the same time, of course, the wanted couplings will have to be forbidden
by appropriate symmetries.

As a specific example, consider a model where three $Z_i\oplus\bar{Z}_i$
fields are present, $i=1,2,3$, acquiring the vevs~(\ref{eq:Z1}),
(\ref{eq:Z2}), (\ref{eq:Z3}) respectively.
Suppose further that the theory has a ${\cal Z}_3^{(1)}\otimes {\cal
Z}_3^{(2)}$
discrete symmetry, under which
\begin{eqnarray*}
{\cal Z}_3^{(1)}&:&\{\bar{H}',H'',Z_1\}\to e^{2\pi i/3}\{\bar{H}',H'',Z_1\},
\qquad\bar{Z}_1\to e^{-2\pi i/3} \bar{Z}_1\\
{\cal Z}_3^{(2)}&:&\{\bar{H}'',H',Z_3\}\to e^{2\pi i/3}\{\bar{H}'',H',Z_3\},
\qquad\bar{Z}_3\to e^{-2\pi i/3} \bar{Z}_3\\
\end{eqnarray*}
These discrete symmetries, without inhibiting the required terms in
$W(Z_i,\bar{Z}_i)$, restrict the renormalizable couplings between the
$H$ and the $Z$-fields in the superpotential to be of the form
$$\bar{H}' Z_1 H'' + H'\bar{Z}_3\bar{H}''.$$
As a result, one obtains a pair of massless Higgs doublets, only contained
in $H',\bar{H}''$.
Even non-renormalizable couplings, if present, can be inhibited up to a
sufficiently high level not to disturb the lightness of these doublets.
On the other hand, it is easy to extend the discrete symmetries to the matter
multiplets, in such a way that the allowed Yukawa couplings be of the
form
$$10'~10'H'+\bar{5}'~10'\bar{Z}_2 \bar{H}''.$$
Notice in particular the necessary asymmetry between the dimensionality
of the operators responsible for the couplings to the light Higgs
doublets of the up-type quarks and of the down-type quarks
(or of the charged leptons) respectively.

\section{}
The model described here can be trivially extended to the
$\SO(10)\otimes\SO(10)$ gauge group.
In this last case, other than the Higgs supermultiplets in the
fundamental vector representations
$$
Z_i=(10,10)_i,\qquad
H'=(10,1),\qquad H''=(1,10),
$$
one must also have the fundamental spinor representation, e.~g.
$$\Psi'=(16,1),\qquad\bar\Psi''=(1,\overline{16})$$
to reduce, as usual, the rank of the group.

In particular, it is easy to see how the discussion of the Higgs superpotential
can be adapted to the SO(10) case.
Experts will recognize a variant of
the Dimopoulos-Wilczek~\cite{DW} mechanism,
designed in SO(10) to undestand the doublet-triplet splitting,
as a natural consequence of the present scheme,
which can in fact be trivially applied to any $\SU(n)\otimes\SU(n)$ group.

In this last case, a model which looks to us especially elegant and
economical is based on the $\SU(6)\otimes\SU(6)$ group.
Suppose that the Higgs supermultiplets only contain the representations
$$\begin{array}{ll}
Z=(6,\bar6),\qquad & \bar{Z}=(\bar6,6)\\
H=(1,\bar6),\qquad & \bar{H}=    (1,6)\end{array}$$
and that the Higgs superpotential is of the form
$$W=W^{(1)}(Z,\bar{Z}) + W^{(2)}(H,\bar{H})$$
with no $Z-H$ interaction, at least up to some level.
Suppose further that $W^{(1)}$ and $W^{(2)}$ be such that, among their minima,
one is obtained for
\begin{eqnarray*}
Z=\bar{Z} &=& V_Z\cdot\diag(1,1,1,1,x,x)\qquad x\neq1\\
H=\bar{H} &=& V_H\cdot(1,0,0,0,0,0)^T
\end{eqnarray*}
This model represents a simpler, maybe string-derivable, variant of a model
previously discussed~\cite{SU66}.
The larger symmetry of the Higgs superpotential, with an extra
global SU(6) factor, leads, after spontaneous symmetry breaking, to the
masslessness of the Higgs doublets whereas the Goldstone triplets are
eaten by the simultaneous breaking of the gauge group.
It has also been noticed elsewhere~\cite{SU6flavour} how this asymmetry
between the SU(2) doublets and the SU(3) triplets in $H,\bar{H}$ may inhibit or
prevent at all the proton decay operators mediated by the colour
triplets.
This comes about because of the absence of any $F$-term-type mass for the
SU(3) triplets.

\section{}
In conclusion, the focus on the successful unification of the gauge
couplings in standard GUTs and on the constraints on model building from
string theory suggests to consider models based on the group $G\otimes G$,
spontaneously broken by Higgs supermultiplets in the fundamental
representation.
It is surprising to see how this viewpoint may lead, at the same time,
to a simple solution of the classic problems of standard GUTs.
We presume that it should be interesting to study possible constraints on
these kinds of models from the point of view of the string theory construction.
{}From our side, we plan to concentrate on the fermion mass problem, along
lines similar to those explored in ref.~\cite{SU6flavour}.

\section*{Acknowledgements}
We thank Zurab Berezhiani and Goran Senjanovi\'c for discussions
on matters related to those ones presented here.

\nonfrenchspacing
\end{document}